\documentstyle[twoside,psfig]{article}

\catcode`\@=11
\long\def\@makefntext#1{
\protect\noindent \hbox to 3.2pt {\hskip-.9pt  
$^{{\eightrm\@thefnmark}}$\hfil}#1\hfill}		%CAN BE USED 

\def\@makefnmark{\hbox to 0pt{$^{\@thefnmark}$\hss}}	%ORIGINAL 
	
\def\ps@myheadings{\let\@mkboth\@gobbletwo
\def\@oddhead{\hbox{}
\rightmark\hfil\eightrm\thepage}   
\def\@oddfoot{}\def\@evenhead{\eightrm\thepage\hfil
\leftmark\hbox{}}\def\@evenfoot{}
\def\sectionmark##1{}\def\subsectionmark##1{}}

\oddsidemargin=\evensidemargin
\newcounter{sectionc}\newcounter{subsectionc}\newcounter{subsubsectionc}
\renewcommand{\section}[1] {\vspace{12pt}\addtocounter{sectionc}{1} 
\setcounter{subsectionc}{0}\setcounter{subsubsectionc}{0}\noindent 
	{\tenbf\thesectionc. #1}\par\vspace{5pt}}
\renewcommand{\subsection}[1] {\vspace{12pt}\addtocounter{subsectionc}{1} 
	\setcounter{subsubsectionc}{0}\noindent 
	{\bf\thesectionc.\thesubsectionc. {\kern1pt \bfit #1}}\par\vspace{5pt}}
\renewcommand{\subsubsection}[1] {\vspace{12pt}\addtocounter{subsubsectionc}{1}
	\noindent{\tenrm\thesectionc.\thesubsectionc.\thesubsubsectionc.
	{\kern1pt \tenit #1}}\par\vspace{5pt}}
\newcommand{\nonumsection}[1] {\vspace{12pt}\noindent{\tenbf #1}
	\par\vspace{5pt}}

\newcounter{appendixc}
\newcounter{subappendixc}[appendixc]
\newcounter{subsubappendixc}[subappendixc]
\renewcommand{\thesubappendixc}{\Alph{appendixc}.\arabic{subappendixc}}
\renewcommand{\thesubsubappendixc}
	{\Alph{appendixc}.\arabic{subappendixc}.\arabic{subsubappendixc}}

\renewcommand{\appendix}[1] {\vspace{12pt}
        \refstepcounter{appendixc}
        \setcounter{figure}{0}
        \setcounter{table}{0}
        \setcounter{lemma}{0}
        \setcounter{theorem}{0}
        \setcounter{corollary}{0}
        \setcounter{definition}{0}
        \setcounter{equation}{0}
        \renewcommand{\thefigure}{\Alph{appendixc}.\arabic{figure}}
        \renewcommand{\thetable}{\Alph{appendixc}.\arabic{table}}
        \renewcommand{\theappendixc}{\Alph{appendixc}}
        \renewcommand{\thelemma}{\Alph{appendixc}.\arabic{lemma}}
        \renewcommand{\thetheorem}{\Alph{appendixc}.\arabic{theorem}}
        \renewcommand{\thedefinition}{\Alph{appendixc}.\arabic{definition}}
        \renewcommand{\thecorollary}{\Alph{appendixc}.\arabic{corollary}}
        \renewcommand{\theequation}{\Alph{appendixc}.\arabic{equation}}
        \noindent{\tenbf Appendix \theappendixc #1}\par\vspace{5pt}}
\newcommand{\subappendix}[1] {\vspace{12pt}
        \refstepcounter{subappendixc}
        \noindent{\bf Appendix \thesubappendixc. {\kern1pt \bfit #1}}
	\par\vspace{5pt}}
\newcommand{\subsubappendix}[1] {\vspace{12pt}
        \refstepcounter{subsubappendixc}
        \noindent{\rm Appendix \thesubsubappendixc. {\kern1pt \tenit #1}}
	\par\vspace{5pt}}

\newcommand{\textlineskip}{\baselineskip=13pt}
\newcommand{\smalllineskip}{\baselineskip=10pt}

\def\eightcirc{
\begin{picture}(0,0)
\put(4.4,1.8){\circle{6.5}}
\end{picture}}
\def\eightcopyright{\eightcirc\kern2.7pt\hbox{\eightrm c}}

\def\abstracts#1#2#3{{
	\centering{\begin{minipage}{4.5in}\baselineskip=10pt\footnotesize
	\parindent=0pt #1\par 
	\parindent=15pt #2\par
	\parindent=15pt #3
	\end{minipage}}\par}} 

\renewenvironment{thebibliography}[1]
	{\frenchspacing
	 \ninerm\baselineskip=11pt
	 \begin{list}{\arabic{enumi}.}
        {\usecounter{enumi}\setlength{\parsep}{0pt}     
	 \setlength{\leftmargin 12.7pt}{\rightmargin 0pt} %FOR 1--9 ITEMS
         \setlength{\itemsep}{0pt} \settowidth
	{\labelwidth}{#1.}\sloppy}}{\end{list}}

\newcounter{itemlistc}
\newcounter{romanlistc}
\newcounter{alphlistc}
\newcounter{arabiclistc}

\newcommand{\fcaption}[1]{
        \refstepcounter{figure}
        \setbox\@tempboxa = \hbox{\footnotesize Fig.~\thefigure. #1}
        \ifdim \wd\@tempboxa > 5in
           {\begin{center}
        \parbox{5in}{\footnotesize\smalllineskip Fig.~\thefigure. #1}
            \end{center}}
        \else
             {\begin{center}
             {\footnotesize Fig.~\thefigure. #1}
              \end{center}}
        \fi}

\newcommand{\tcaption}[1]{
        \refstepcounter{table}
        \setbox\@tempboxa = \hbox{\footnotesize Table~\thetable. #1}
        \ifdim \wd\@tempboxa > 5in
           {\begin{center}
        \parbox{5in}{\footnotesize\smalllineskip Table~\thetable. #1}
            \end{center}}
        \else
             {\begin{center}
             {\footnotesize Table~\thetable. #1}
              \end{center}}
        \fi}

\def\@citex[#1]#2{\if@filesw\immediate\write\@auxout
	{\string\citation{#2}}\fi
\def\@citea{}\@cite{\@for\@citeb:=#2\do
	{\@citea\def\@citea{,}\@ifundefined
	{b@\@citeb}{{\bf ?}\@warning
	{Citation `\@citeb' on page \thepage \space undefined}}
	{\csname b@\@citeb\endcsname}}}{#1}}

\newif\if@cghi
\def\cite{\@cghitrue\@ifnextchar [{\@tempswatrue
	\@citex}{\@tempswafalse\@citex[]}}
\def\citelow{\@cghifalse\@ifnextchar [{\@tempswatrue
	\@citex}{\@tempswafalse\@citex[]}}
\def\@cite#1#2{{$\null^{#1}$\if@tempswa\typeout
	{IJCGA warning: optional citation argument 
	ignored: `#2'} \fi}}

\def\pmb#1{\setbox0=\hbox{#1}
	\kern-.025em\copy0\kern-\wd0
	\kern.05em\copy0\kern-\wd0
	\kern-.025em\raise.0433em\box0}

\def\fnt#1#2{\footnotetext{\kern-.3em
	{$^{\mbox{\scriptsize #1}}$}{#2}}}

\def\fpage#1{\begingroup
\voffset=.3in
\thispagestyle{empty}\begin{table}[b]\centerline{\footnotesize #1}
	\end{table}\endgroup}

\def\runninghead#1#2{\pagestyle{myheadings}
\markboth{{\protect\footnotesize\it{\quad #1}}\hfill}
{\hfill{\protect\footnotesize\it{#2\quad}}}}
\font\tenrm=cmr10
\font\tenit=cmti10 
\font\tenbf=cmbx10
\font\bfit=cmbxti10 at 10pt
\font\ninerm=cmr9

\font\eightrm=cmr8

\def\qed{\hbox{${\vcenter{\vbox{			%HOLLOW SQUARE
   \hrule height 0.4pt\hbox{\vrule width 0.4pt height 6pt
   \kern5pt\vrule width 0.4pt}\hrule height 0.4pt}}}$}}

\begin{document}

\runninghead{Orbiting valence quarks and their influence on
$\ldots$} {Orbiting valence quarks and their influence on $\ldots$}

\normalsize\textlineskip
\thispagestyle{empty}
\setcounter{page}{1}

%\copyrightheading{}			%{Vol. 0, No.0 (1992) 000--000}

%\vspace*{0.88truein}

\fpage{1}
\centerline{\bf ORBITING VALENCE QUARKS AND THEIR INFLUENCE ON}
\vspace*{0.035truein}
\centerline{\bf THE STRUCTURE FUNCTIONS OF THE NUCLEON}
\vspace*{0.37truein}
\centerline{\footnotesize LIANG ZUO-TANG$^{\dag\ddag}$ 
and R. RITTEL$^{\dag}$}
\vspace*{0.015truein}
\centerline{\footnotesize\it 
$^{\dag}$Institut f\"ur theoretische Physik, Freie Universit\"at Berlin}
\baselineskip=10pt
\centerline{\footnotesize\it Arnimallee 14, 14195 Berlin, Germany}
\vspace*{0.015truein}
\centerline{\footnotesize\it 
$^{\ddag}$Department of Physics, Shandong University}
\baselineskip=10pt
\centerline{\footnotesize\it Jinan 250100, China}

%\vspace*{10pt}
%\centerline{\footnotesize SECOND AUTHOR}
%\vspace*{0.015truein}
%\centerline{\footnotesize\it Group, Laboratory, Address}
%\baselineskip=10pt
%\centerline{\footnotesize\it City, State ZIP/Zone, Country}
\vspace*{0.225truein}
%\publisher{(received date)}{(revised date)}
%\begin{center}                         % FUER DIE EIGEREICHTE
%  \footnotesize\smalllineskip          % (SUBMITTED) VERSION
%   Submitted (November 25, 1996)
%\end{center}

\vspace*{0.21truein}
\abstracts{It is shown that 
intrinsic orbital motion of the 
valence quarks has large influences on 
the spin-dependent as well as 
the spin-averaged nucleon structure functions. 
Its connection with the 
observed ``very small contribution of quark spin 
to nucleon spin'' and the observed violation 
of Gottfried sum rule is discussed.}{}{}

%\vspace*{10pt}
%\keywords{The contents of the keywords}

\textlineskip			%) USE THIS MEASUREMENT WHEN THERE IS
\vspace*{12pt}			%) NO SECTION HEADING

%\vspace*{1pt}\textlineskip	%) USE THIS MEASUREMENT WHEN THERE IS
%\section{General Appearance}	%) A SECTION HEADING
%\vspace*{-0.5pt}
\noindent
Deep inelastic lepton-hadron scattering 
plays a rather unique role
in studying the structure of hadron.
Through measurements of the structure functions 
in such experiments, people have learned$^{1}$ that 
hadrons are made out of 
point-like constituents --- quarks. 
Based on the picture 
of the quark parton model$^{2}$, 
a set of relations have been derived 
between the structure functions and the 
quark distributions 
in the infinite momentum frame, 
and a rather simple interpretation 
of the Bjorken variable in that frame 
has been obtained. 
Parton model itself says nothing about 
the forms of the structure functions 
but the integrals of them. 
It is predicted$^{3}$ that a set of sum rules 
should be valid. 
Such sum rules link the structure functions to 
quantities which can be measured 
in other kinds of experiments 
such as those associated with 
the static properties of the hadrons.
Actually, the sum rules are the only places where 
structure functions and static properties
of hadrons meet each other in the parton model 
and it is also the sum rules 
which can be checked experimentally.
With increasing accuracy of 
the measurements, large discrepancies 
have been observed$^{4,5,6,7,8}$ 
between data and theory.
The two most well-known examples are the integral of 
the spin-dependent structure function 
$g_1(x_B)$ and the Gottfried sum. 
It is found$^{4-8}$ that 
both of them are much smaller 
than those expected in the model.
The former led$^{4,5,6,7}$ to the conclusion that 
quark spin contributes only a very small fraction 
to the nucleon spin and 
thus triggered the ``spin crisis''. 
The latter raised$^{9}$ the question 
whether isospin-invariance is violated 
in the nucleon sea.

In a series of papers published recently, it has been
pointed out$^{10}$ 
that intrinsic motion of the confined quarks plays 
a very important role in understanding 
the polarization phenomena in high energy 
collisions, and that relativistic quark models
can be constructed which reproduce baryon's magnetic 
moments on the one hand, and
describe the observed$^{11}$ 
left-right asymmetries in inclusive 
meson or hyperon production 
in high energy processes on the other.  
It has been shown$^{10}$ in particular that 
once we accept that quarks are spin-$1/2$ 
particles moving in a confined 
spatial region, we are forced also to accept 
that orbital motion of such valence quarks is 
always involved, even when they are 
in their ground states.
In other words, in relativistic quark models, 
intrinsic motion of the valence quarks 
appears simply as orbital motion. 
The average orbital angular momenta of 
the valence quarks are simply nonzero 
if the nucleon is polarized.
In this connection it is also 
interesting to see that
orbital angular momenta of quarks were 
expected$^{12}$ to contribute to the 
proton spin by analyzing different data 
in the framework of the parton model and others.
We recall that intrinsic transverse motion 
was neglected in the formulation of 
the parton model$^{2}$,   
and now it is usually thought that 
transverse motion contributes only to high twist 
effects which vanish at high $Q^2$.
We are therefore led to the following questions:
What kind of effects do such orbital motions have 
on the structure functions of the nucleon?
Can we understand the above mentioned data if we take 
them into account? 
Are the above mentioned effects 
observed in polarized$^{4,5,6,7,11}$ and 
unpolarized$^{8}$ experiments
connected with each other? 
These are questions we would like to 
discuss in this note. 
We ask in this connection also: Should not these questions 
be made clear before we seek for other dynamical origins 
of the above mentioned effects observed experimentally?

Since orbital motion of 
the valence quarks are best described in 
nucleon's rest frame, we will discuss these 
questions also in that frame. 
We consider the inclusive process $l+N\to l\ +$ anything,
where $l$ stands for electron or muon, $N$ for nucleon, 
and denote the 4-momentum of $N$ 
and those of the incident and the scattered $l$'s 
by $P=(M,\vec 0),\ k$, $k'$ and 
the 4-momentum transfer 
carried by the virtual photon $\gamma^*$ by 
$q \equiv (\nu, \vec q)=k-k'$ respectively.
We study deep inelastic scattering processes  
in the Bjorken limit, i.e. $Q^2\equiv -q^2$ 
is very large ($Q^2\gg M^2$) 
while $x_B \equiv Q^2/(2M\nu )$, 
the Bjorken-$x$, is kept fixed.
For such events, it is observed that$^{1}$ 
Bjorken scaling is approximately valid, 
and thus it is expected that 
the Bjorken variable $x_B$ 
should play a special role 
in describing such events. 
Hence, the questions we immediately encounter are: 
What does $x_B$ mean 
in the rest frame of the nucleon? 
Why is it particularly useful in 
describing deep inelastic scattering events?
What does the existence of the 
approximate Bjorken scaling tell us about 
the structure of nucleon in its rest frame?
We recall that in the parton model$^{2}$, 
one treats the problem in the 
infinite momentum frame where 
the transverse motion of the quarks 
and the interactions between them 
during the lepton-nucleon interaction 
can be neglected. 
Because of this, one 
obtains a very simple 
interpretation of $x_B$. 
It is simply the fractional (longitudinal) 
momentum of the struck quark with respect to 
the nucleon in that frame.
How is the situation in nucleon's rest frame 
--- what is still applicable 
and what is no more valid here?
To answer these questions we 
note the following: 

(i) In deep inelastic scattering, 
$Q^2$ is very large. 
This implies a high spatial resolution 
so that the interaction between the lepton and 
the quark is point-like. 
The time interval $\Delta t$, in which 
such a point-like interaction takes place, 
is proportional to $1/\nu$. 
In the Bjorken limit, 
$\Delta t$ is much shorter than the typical time 
needed for color propagation 
between the quarks in the nucleon. 
The interaction 
between the lepton and the confined quark 
is already over, before the latter 
could exchange energy-momentum or 
any other quantum 
numbers with the neighboring quarks. 
In other words, 
{\it impulse approximation$^{13}$ is valid 
for such scattering processes, 
also in nucleon' rest frame}. 
The confining potentials determine 
the initial states of the quarks but 
can be neglected during the 
lepton-quark interaction. 
This is to be compared with the parton model in the 
infinite momentum frame, 
where not only the interactions between the quarks 
are neglected but also the initial states 
of the quarks are taken as plane waves 
with momenta in the same direction as the nucleon.

(ii) At large $Q^2$ and 
fixed $x_B\equiv Q^2/(2M\nu )$, we have, 
$\Delta \equiv |\vec q \ |-\nu 
\approx Q^2/(2\nu)$,
so that 
\begin{equation}
x_B\equiv Q^2/(2M\nu )\approx  \Delta / M.
\end{equation}
This means that the
virtual photon $\gamma^*$ has 
an energy-deficit$^{14}$, 
and that {\it $x_B$ is 
nothing else but the energy-deficit of the 
virtual photon in unit of proton mass.}   

(iii) During the lepton-quark interaction, 
the virtual photon $\gamma ^*$ is absorbed 
by the quark inside the nucleon.
The cross sections or 
the structure functions derived from them 
are proportional to the probability 
for such an absorption to take place.

The absorption of $\gamma^*$ by a
confined point-like quark implies that  
the latter obtains, 
not only an enormously large amount of momentum, 
but also the corresponding energy and thus 
the above mentioned energy-deficit.
The initial 4-momentum 
$p\equiv (\varepsilon ,
p_{\parallel }, \vec p_\perp)$ of the 
struck quark is suddenly and drastically changed to  
$p' \equiv (\varepsilon',p'_{\|},\vec p_\perp )=p+q$ 
(where $_\parallel $ is defined wrt the direction of $\vec q$).  
As a consequence, it moves kinematically like a free particle 
such that a current jet (of hadrons) can be produced. 
The necessary and sufficient condition for this 
to occur is ${p'}^2=m_q^2$, 
which leads$^{15}$ to 
$ \varepsilon -p_{\|} \approx \Delta $ in the Bjorken limit.  
This implies that the struck quark should 
have an ``energy excess'' 
and this ``energy excess'' 
approximately compensates the energy deficit $\Delta $ 
of the virtual photon $\gamma ^*$. 
%so that latter can be absorbed by the former to form a current jet. 
In terms of $x_B$, this condition is:
\begin{equation}
\varepsilon-p_{\|} \approx \Delta \approx Mx_B. 
\end{equation}
That is to say:
Among all the (point-like and confined) 
quarks in the target nucleon, 
the virtual photon may encounter 
various quarks in different states, 
but only those which have the 
right ``energy-excess'' 
at the moment when they get struck 
contribute to the observed current jet.
The virtual photon $\gamma ^*$ 
can have different momentum $\vec q$ and energy $\nu $, 
but for the deep inelastic scattering 
to take place, only {\it its energy deficit $\Delta $ or 
the quantity $x_B\approx \Delta /M$ 
is relevant}. This shows why $x_B$ is 
particularly useful in describing these events.

(iv) The cross section for deep inelastic scattering 
at fixed $x_B$ is therefore 
determined by the probability for finding 
quarks with ``energy excess'' 
$\varepsilon -p_\| \approx \Delta \approx x_BM$ in the nucleon. 
It is clear that if $Q^2$ is already large enough, 
i.e. the spatial resolution is high enough so that 
the point-like constituents can already be resolved,
further increasing of $Q^2$, 
which means further increasing of spatial resolution, 
will not see anything new. 
This implies that the probability 
for finding such quarks 
should be independent of $Q^2$ and thus 
the Bjorken scaling is valid in this case.

Having seen that the qualitative 
features of deep inelastic scattering 
can indeed also be understood without introducing 
the infinite momentum frame, we now study 
the influence of 
the orbital motion of the valence quarks on 
the (spin-dependent as well as spin averaged) 
structure functions. 
We treat this problem in 
the rest frame of the nucleon,  and  
recall$^{10}$ the following: 
In this frame, the valence quark $q_v$ 
can and should be 
described by the spherical wave which 
is an eigenstate of  
four operators: the Hamiltonian $\hat H$, 
the total angular momentum squared $\hat {{\vec j}^2}$, 
its third component $\hat j_z$,
and the parity $\hat {\cal P}$ 
with eigenvalues $\varepsilon $, $j(j+1)$, $m$ 
and $\cal {P}$ respectively.   
In momentum space, it is given by$^{/16/}$, 
\begin{equation}
\tilde \psi_{\varepsilon j m{\cal P}} (\vec p\ |q_v)
=(-i)^\ell \left( \matrix{ 
\tilde f_{\varepsilon\ell} (p|q_v)
\Omega^{jm}_{\ell } (\theta,\phi)\cr 
-i\tilde g_{\varepsilon\ell'} (p|q_v)\Omega^{jm}_{\ell'} 
(\theta,\phi)\cr} \right), 
\end{equation}
where ${\cal P}=(-1)^l, j=l\pm 1/2$ and $l'=l\pm 1$. 
Here, as well as in the following, $p$, 
when used as argument 
in $\tilde f$ or $\tilde g$,  
stands for $|\vec p \ |$. 
We see clearly that orbital motion 
is always involved even if the 
quark $q_v$ is in its ground state 
$\tilde \psi_{\varepsilon j m{\cal P}} $ 
where $\varepsilon =\varepsilon _0, 
j=1/2, m=\pm 1/2 $ and ${\cal P}=+$ 
(i.e. $l=0,l'=1)$.
Hence, we consider, as the first step, 
the following demonstrating example: 
We assume that the valence quarks can be 
treated, just like that 
in the quark parton model,  
as free but they are in the above 
mentioned eigenstates of $\hat H$, 
$\hat {{\vec j}^2}$, $\hat j_z$
and $\hat {\cal P}$. 
The two radial functions $\tilde f$ and $\tilde g$ 
are determined by the Dirac equation for 
free particle.  
We calculate the contribution of 
one of such quarks, $q_v$,  
to the structure functions of the nucleon, 
and compare the results with those obtained 
in the parton model.
We recall that the S-matrix element 
for the elementary process 
$e^-q_v\to e^-q_v$ is given by,
\begin{eqnarray}
S^{e^-q_v}_{fi} & = & -i\int d^4x\int d^4y 
\left\{ \Bigl [ -e\Psi^{(e)\dagger }_f(x)
\gamma _\alpha \Psi ^{(e)}_i(x) \Bigr ] \right. \nonumber\\
 & & \hspace*{3cm}\times
\left. D_F(x-y) 
\Bigl [ ee_{qv}\Psi^{(qv)\dagger }_f(y)
\gamma ^\alpha \Psi ^{(qv)}_i(y) \Bigr ] \right\} .
\end{eqnarray}
Here $D_F(x-y)=\int d^4q (-1/q^2) e^{-iq(x-y)}/(2\pi )^4$ 
is the photon propagator, 
the $\Psi $'s are 
the initial and the final 
(denoted by the subscripts $i$ and $f$ respectively) 
state wave functions for the electron 
and the quark 
[denoted by the superscripts $(e)$ and $(qv)$ respectively] 
in coordinate space. 
They are chosen as follows: 
The initial and final states 
for the electron are 
plane waves with 4-momentum $k$ and $k'$ respectively.
The final state for the quark is 
plane wave with 4-momentum $p'$ but 
the initial state is the spherical wave given by Eq.(3).
We insert them into Eq.(4) and obtain the contribution of 
this elementary process to the hadronic tensor 
$W^{\alpha \beta}(P,S;q)$ 
(where $S$ stands for the polarization 
of the nucleon). 
Its contribution to the structure functions can 
then be calculated in a straight forward manner. 
The results obtained 
in the Bjorken limit
for a $q_v$ in its ground state  
$\tilde \psi_{\varepsilon_0 \ {1\over 2} \ m\  +}$ 
are given by,
\begin{eqnarray}
F_{2(qv)}(x_B|m) & \approx & {Mx_Be^2_{qv} \over 2}
\int p_\perp dp_\perp 
\bigl [\tilde f_{00}^2 (p|q_v)+ \tilde g_{01}^2 (p|q_v) +\nonumber\\
& &
\hspace*{3cm}
2\cos \theta \tilde f_{00} (p|q_v) 
\tilde g_{01}(p|q_v)\bigr ];
\end{eqnarray}
\begin{eqnarray}
g_{1(qv)}(x_B|m) & \approx & m{Me_{qv}^2\over 2}
\int p_\perp dp_\perp 
\bigl [\tilde f_{00}^2 (p|q_v) +(1-2\sin^2\theta ) 
\tilde g_{01}^2 (p|q_v) +\nonumber\\
 & & \hspace*{3cm}
2\cos\theta \tilde f_{00} (p|q_v)\tilde g_{01}(p|q_v)\bigr ];\\ 
g_{2(qv)}(x_B|m) & \approx & m{Me_{qv}^2\over 2} 
\int p_\perp dp_\perp 
\bigl [ (1-3 \cos^2\theta) \tilde g_{01}^2 (p|q_v) - \nonumber\\
 & & \hspace*{3cm} 
2\cos\theta \tilde f_{00}(p|q_v)\tilde g_{01}(p|q_v)\bigr ], 
\end{eqnarray}
and $F_{1(qv)}(x_B|m)
\approx F_{2(qv)}(x_B|m)/(2x_B)$.
Here $\tilde f_{00}(p|q_v) \equiv 
\tilde f_{\varepsilon\ell} (p|q_v)$ 
for $\varepsilon=\varepsilon_0,\ell=0$
and $\tilde g_{01}(p|q_v)\equiv 
\tilde g_{\varepsilon\ell'} (p|q_v)$ for
$\varepsilon=\varepsilon_0,\ell '=1$; 
$\cos\theta\equiv p_\parallel/p $,
and $p_\parallel \approx \varepsilon_0 - Mx_B$. 
The integration over $p_\perp $ is carried in the 
region as given in [15].

These results are interesting since they show in particular the following:

(A) From Eq.(5), we see: 
$F_{2(qv)}(x_B|m)$ 
contains not only terms proportional to 
the quark density
$|\tilde \psi (\vec p\ )|^2\propto 
\tilde f_{00}^2(p)+\tilde g_{01}^2(p)$ 
but also the ``mixed term'' 
$\cos \theta \tilde f_{00}(p)\tilde g_{01}(p)$. 
Hence, the nucleon structure function $F_2$ 
is {\it not} just proportional to the number densities 
of quarks in the nucleon. 
This is essentially different from that in the 
quark parton model.

(B) For $g_{1(qv)}(x_B|m)$,
the integrand contains, besides 
terms like $|\tilde \psi (\vec p\ )|^2$ 
and the ``mixed term", an additional term 
$-2\sin^2 \theta \tilde g_{01}^2(p)$   
which is negative in sign and 
is proportional to $\tilde g_{01}^2(p)$. 
This is expected 
because $\tilde g_{01}^2(p)$ 
comes from the lower component of $\tilde \psi $ 
and such a component corresponds to $l'=1$. 
Its contribution to the 
spin-dependent structure functions 
should be different from 
the upper component which corresponds to $l=0$.
It should be emphasized that,  
in contrast to the usual expectations, 
{\it neither of these terms vanishes 
even in the limit $Q^2\to \infty$.} 

(C) Since $x_B\approx 
(\varepsilon _0-p_\parallel )/M$, 
the interval $0\le x_B\le 1$ 
corresponds to $\varepsilon _0 \ge 
p_\| \ge (\varepsilon _0-M)$. 
This is in general {\it not} necessary the {\it entire} 
physical region for the momenta 
of the bound valence quarks. 
Hence, the integral over this range is {\it not} 
the sum over all possible states of the bound quarks!
In particular, by integrating 
$F_{2(qv)}(x_B|m)/x_B$ 
over $x_B$ from zero to unity, 
we do not get $e_{qv}^2$ but 
a number which is in general less than it. 
It tends to $e_{qv}^2$ in the static limit, 
where we have $|p_\| |\ll M$. 
This means, sum rules such as 
those in the parton model 
are in general {\it not} valid here. 
The results of such 
integrals should be, in most of the cases, 
less than those expected in the parton model. 
This implies, e.g.,  
the Gottfried sum$^{3}$, which is the 
integral of $[F_{2}^p(x_B)-F_{2}^n(x_B)]/x_B$, 
should be less than $1/3$. 
It is $1/3$ only in the 
static limit$^{17}$.  
But, in this limit, 
the integrand, namely the structure functions, 
will have the form of a Delta-function
--- a distribution  
which contradicts the existing data$^{1,8,18,19,20}$.

(D) Not only because of 
the facts pointed out in (C) 
but also due to the 
presence of the term $-2\sin ^2\theta \tilde g_{01}^2(p)$, 
the integration of 
$g^{p}_{1(q_v)}(x_B)$, and thus that of
$g_1^p(x_B)$ over $x_B$ from zero to one is 
expected to be much smaller than that expected in the 
quark parton model. 
This is consistent with the 
experimental observations$^{4-8}$.

(E) Similar discussions as those given in (C) 
and (D) show 
that, strictly speaking,
just as Gottfried sum rule,  
Bjorken sum rule$^{3}$ should also be violated.  
However, if we compare these two sum rules, 
we see the following difference:
While both sides of Bjorken sum rule 
depend on the radial wave functions$^{21}$, 
the rhs of Gottfried sum rule does not.
Hence, in the relativistic case, 
both sides of Bjorken sum rule 
and the lhs of Gottfried sum rule should be 
much smaller than their counterparts 
in the static limit, 
while the rhs of Gottfried sum rule 
remains the same. 
Since these sum rules are valid in the static limit,  
this implies a strong violation of Gottfried sum rule,  
but only a weak violation of Bjorken sum rule 
in the relativistic models. The latter can even be approximately valid
for some particular choices of $\tilde{f}_{00}$ and
$\tilde{g}_{01}$. Also this is consistent with the data$^{5,6}$.

Encouraged by these agreements, we
consider a valence quark
in the mean field 
caused by the other constituents of the nucleon. 
We take the mean field as central and 
describe the valence quark by 
the spherical wave given by Eq.(3) 
in this central field.  
The calculations of the contributions 
of these valence quarks to the structure 
of the nucleon can be carried out 
in exactly the same way as above.
For one quark, the results have
exactly the same form as those given 
in Eqs.(5)---(7). The only difference is that 
now the radial functions $\tilde f$ and $\tilde g$ 
are solutions of the Dirac equation with given potentials. 
To see how the quantitative 
results from the conventional 
simplest potentials are compared to data,   
we considered
a simple spherical potential well, i.e. 
$U_S(r)=0, U_V=-0.3M$ for $0\le r\le R$ but $U_S(r)=\infty $ for $r>R$,
and obtained the contribution of a valence quark to the
structure functions from Eqs. (5)\,--\,(7).
The contributions of all the valence quarks are then obtained
by summing over all of them, i.e.
\begin{equation}
F_2(x_B)
=\sum_{q_v,m} \rho_0(m;q_v|\rightarrow )
F_{2(qv)}(x_B|m),
\end{equation}
\begin{equation}
g_{1,2}(x_B)=\sum_{q_v,m} \rho_0(m;q_v|\rightarrow )
g_{1,2(qv)}(x_B|m).
\end{equation}
Here, $\rho_0(m,q_v|\rightarrow )$ 
is the average number of
valence quarks in the state 
$\tilde \psi_{\varepsilon_0 \ {1\over 2} \ m\  +}$ 
in the nucleon which 
is polarized in $z$-direction, it
is determined$^{10}$ by the 
nucleon wave function.
We calculated first $F_{2}^p(x_B)-F_{2}^n(x_B)$, 
which is of particular interest, 
not only because it is nothing else but the 
integrand of the Gottfried sum,  
but also because it contains only valence quark contributions 
provided that isospin invariance is not violated 
in the quark-antiquark sea. 
The result is shown in Fig.1.
\begin{figure}[htbp]
\vspace*{13pt}
\centerline{\vbox{\hrule width 5cm height0.0pt}}  %height0.001pt}}
\vspace*{-0.8cm}
\begin{center}
\psfig{figure=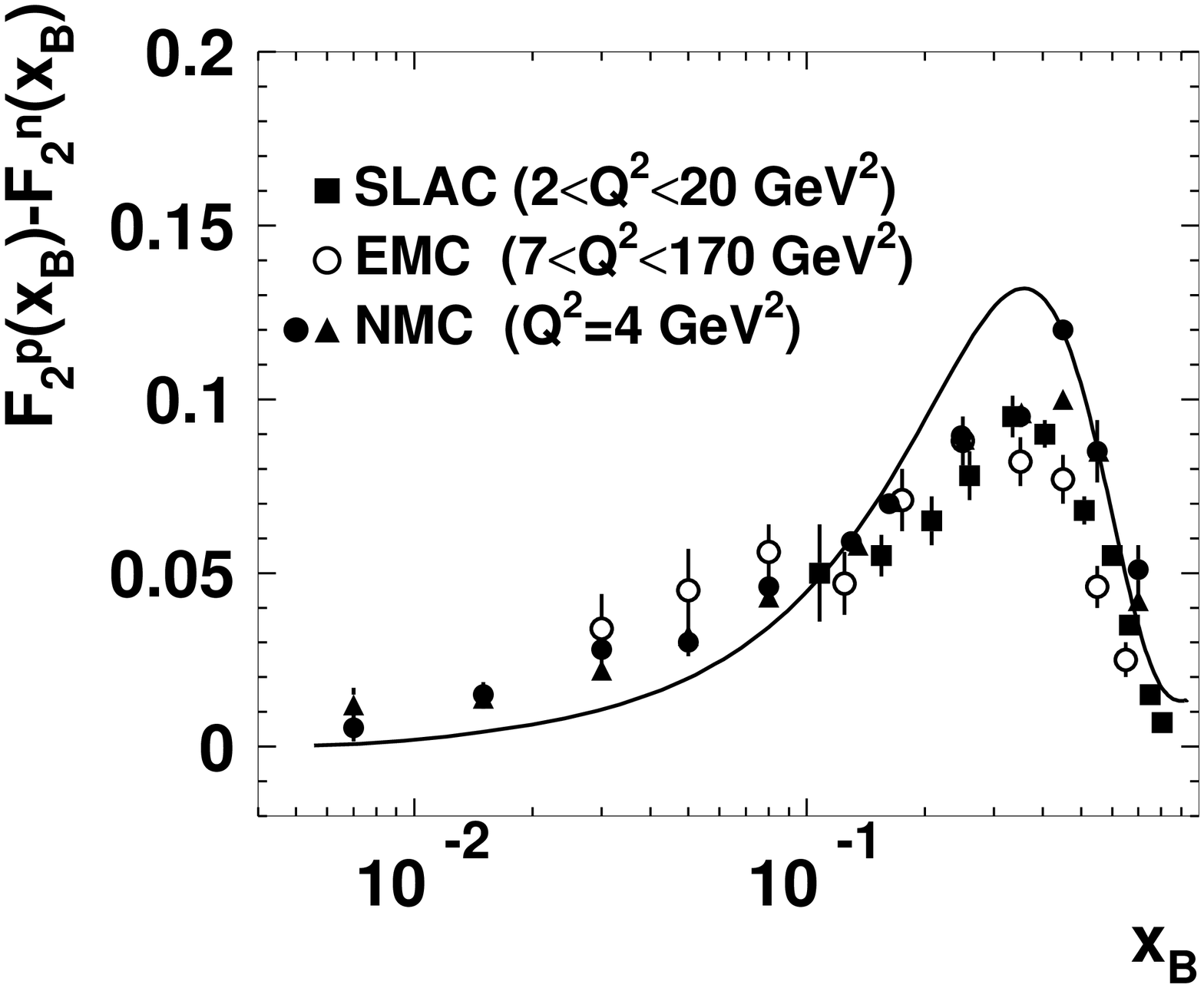,width=10cm}
\end{center}
%\vspace*{1.4truein}		%ORIGINAL SIZE=1.6TRUEIN x 100% - 0.2TRUEIN
\vspace*{-1.8cm}
\centerline{\vbox{\hrule width 5cm height0.0pt}}  %height0.001pt}}
\vspace*{13pt}
\fcaption{
The difference $F_2^p(x_B)-F_2^n(x_B)$ 
as function of $x_B$.  
The curve is the result of Eqs.(8) and (5)
obtained by using 
the $f_{00}(p|q_v)$ and $g_{01}(p|q_v)$ 
from the potential well mentioned in the text 
with $R=1.23$\,fm for both $u$ and $d$.  
Data are taken from [8,18,19] and [20].
(Only statistical errors are shown.)}
\end{figure}
The same solutions have also 
been used to calculate 
$g_1^p(x_B)$ and $g_2^p(x_B)$.
The results are shown in Figs.2 and 3
respectively. 
\begin{figure}[htbp]
\vspace*{13pt}
\centerline{\vbox{\hrule width 5cm height0.0pt}}  %height0.001pt}}
\vspace*{-0.8cm}
\begin{center}
\psfig{figure=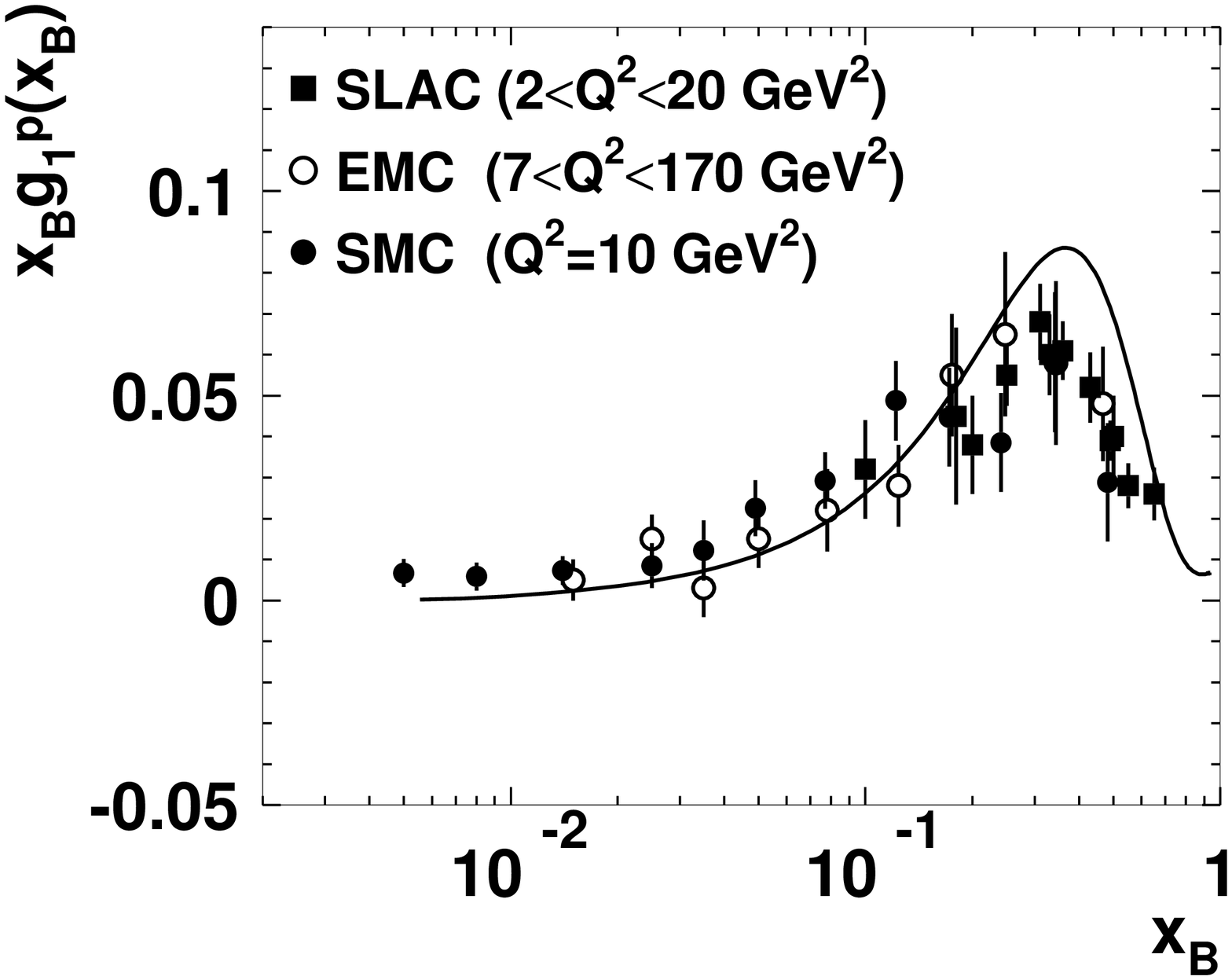,width=10cm}
\end{center}
%\vspace*{1.4truein}		%ORIGINAL SIZE=1.6TRUEIN x 100% - 0.2TRUEIN
\vspace*{-1.8cm}
\centerline{\vbox{\hrule width 5cm height0.0pt}}  %height0.001pt}}
\vspace*{13pt}
\fcaption{
The spin-dependent structure function 
$x_B g_1^p(x_B)$ as function of $x_B$. 
The curve is the result of Eqs.(9) and (6) 
by using the same sets of 
$f_{00}(p|q_v)$ and $g_{01}(p|q_v)$ 
as those in Fig.1. 
The data are taken from Refs.[4,6,7,22] and [23].
(Only statistical errors are shown.)}
\end{figure}
\begin{figure}[htbp]
\vspace*{13pt}
\centerline{\vbox{\hrule width 5cm height0.0pt}}  %height0.001pt}}
\vspace*{-0.8cm}
\begin{center}
\psfig{figure=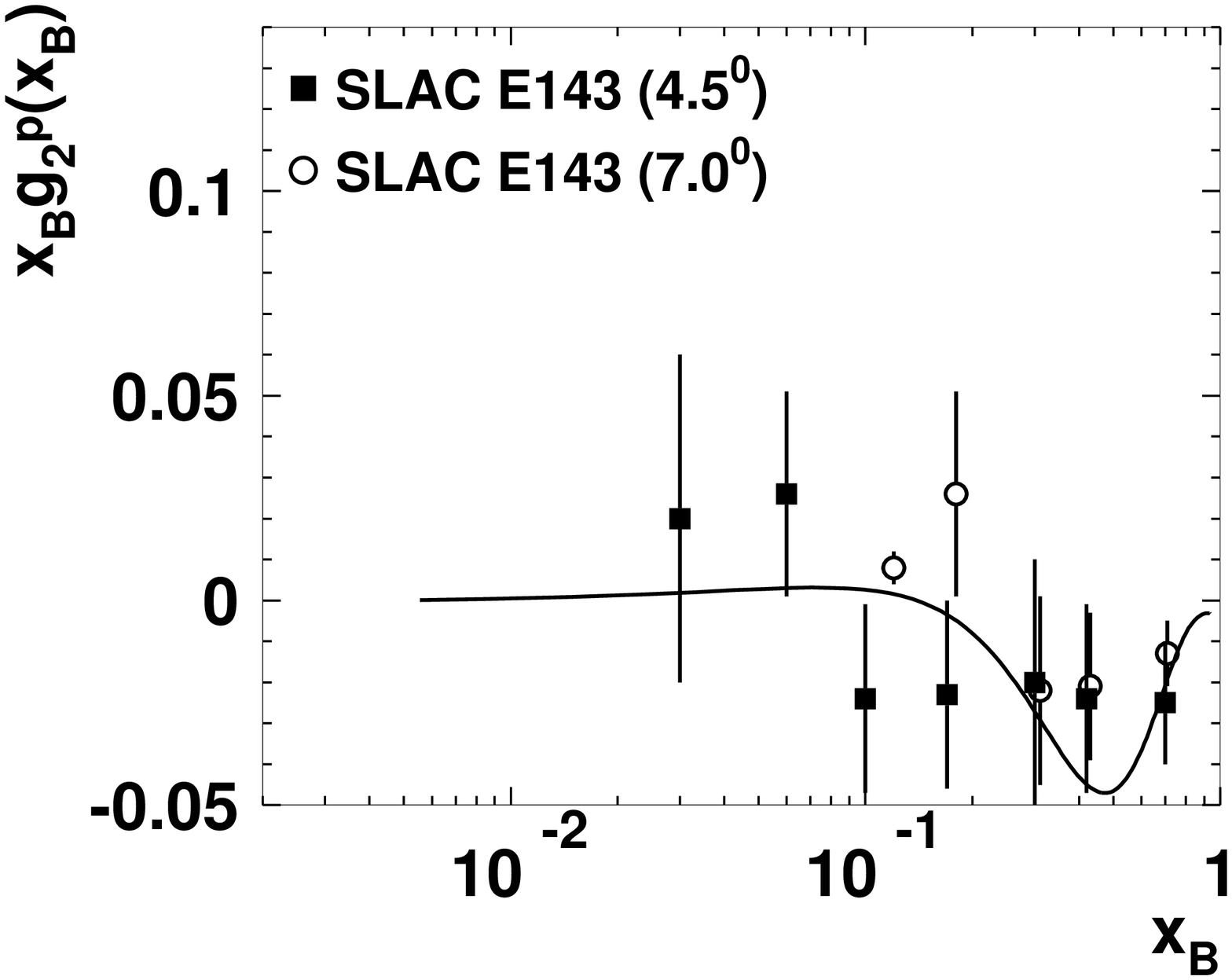,width=10cm}
\end{center}
%\vspace*{1.4truein}		%ORIGINAL SIZE=1.6TRUEIN x 100% - 0.2TRUEIN
\vspace*{-1.8cm}
\centerline{\vbox{\hrule width 5cm height0.0pt}}  %height0.001pt}}
\vspace*{13pt}
\fcaption{The spin-dependent structure function 
$x_B g_2^p(x_B)$ as function of $x_B$. 
The curves are the results obtained from Eqs.(9) and (7) 
by using the same sets of 
$f_{00}(p|q_v)$ and $g_{01}(p|q_v)$ 
as those used in Figs.1 and 2.
The data are taken from [7]. 
(Only statistical errors are shown.)}
\end{figure}

It should be emphasized in this connection that 
our purpose here is to investigate 
the influences of the intrinsic orbital motion 
of valence quarks on the structure functions of nucleon. 
No attempt has yet been made to get a better 
fit to the data 
by making a more suitable choice of 
parameters for the confining potentials, 
although such a procedure is clearly possible.
(We give therefore also no quantitative predictions 
for the integrals of the structure functions since 
such quantitative results depend very much on 
the explicit forms of the confining potentials.)
No difference between the effective potentials 
for $u$- and $d$-valence quarks 
in the nucleon has been taken into account yet. 
We get therefore $g_1^n(x_B)=g_2^n(x_B)=0$.
This shows that the magnitudes of these two
structure functions should be much smaller than those
of their counterparts for the proton 
(i.e. $g_1^p$ and $g_2^p$),
which is consistent with the recent experimental findings.
Non-zero values of $g_1^n$ and $g_2^n$ may for example originate from 
the existence$^{24}$ of the differences between
the wave functions of $u$- and $d$-quarks and/or other
effects, which are not discussed here.
  
In summary, together with illustrative examples, 
we have explicitly demonstrated that 
the intrinsic orbital motion 
of the valence quarks can have profound influence on the 
structure functions of the nucleon. 
The obtained result shows that 
the violation of the sum rules 
derived in the parton model 
is in fact not surprising and that 
the conventional interpretation 
of the nucleon structure functions 
may not be the most useful one. 
This is particularly obvious 
in connection with the spin structure 
of the nucleon.

\nonumsection{Acknowledgements}
\noindent
We thank C. Boros and Meng Ta-chung for helpful discussions.  
This work is supported 
in part by Deutsche Forschungsgemeinschaft (DFG:Me 470/7-1).

\newpage

\nonumsection{References}


\begin{thebibliography}{000}
\bibitem {15} The exact result we obtain is: 
$\varepsilon -p_{\|} =\Delta +(\varepsilon+\nu)\cdot
[1-\sqrt{1-(p_\perp^2+m_q^2)/(\varepsilon+\nu)^2}]$ and 
a limit for $p_\perp $: $0\le p_\perp \le \sqrt {(\varepsilon +\nu)^2-m_q^2}$.
\bibitem {16} The notations used here are the same as  
those in the second paper of Ref.10.
\end{thebibliography}
\end{document}